\documentclass[11pt]{article}

\usepackage[totalwidth=480pt, totalheight=680pt]{geometry}
\usepackage{cite}
\usepackage{mathptmx}
\usepackage{amsmath,amsfonts,amsthm,amsbsy,amssymb}
\usepackage[hidelinks]{hyperref}
\usepackage{mathtools}
\usepackage{nicefrac}
\usepackage{mathrsfs}
\usepackage{bbm}
\numberwithin{equation}{section}
\usepackage{lmodern}
\usepackage[bottom]{footmisc}

\newcommand{\tr}{\mathrm{Tr}}

\begin{document}
\begin{titlepage}
\begin{center}
\vspace*{1.5cm}
\textbf{\LARGE Two Loop Ghost free Quantisation of Wilson Loops in} \\[2ex]
\textbf{\LARGE $\mathcal{N}=4$ supersymmetric Yang-Mills} \\
\vspace{2cm}
\textsc{\Large Hannes Malcha} \\
\vspace{1cm}
\textit{Max-Planck-Institut f\"{u}r Gravitationsphysik (Albert-Einstein-Institut)} \\
\textit{Am M\"{u}hlenberg 1, 14476 Potsdam, Germany} \\
\vspace{.5cm}
{\ttfamily \href{mailto:hannes.malcha@aei.mpg.de}{hannes.malcha@aei.mpg.de}} \\
\vspace{2cm}
\textbf{Abstract}
\end{center}
\small{
\noindent
We report a perturbative calculation of the expectation value of the infinite straight line Maldacena-Wilson loop in $\mathcal{N}=4$ supersymmetric Yang-Mills theory to order $g^6$. Thus, we extend the previous perturbative result by one order.  The vacuum expectation value is reformulated in terms of a non-linear and non-local transformation, the Nicolai map, mapping the full functional measure of the interacting theory to that of a free bosonic theory. The results are twofold. The perturbative cancellations of the different contributions to the Maldacena-Wilson loop are by no means trivial and seem to resemble those of the circular Maldacena-Wilson loop at order $g^4$. Furthermore, we argue that our approach to computing quantum correlation functions is competitive with more standard diagrammatic techniques.
}
\vspace{\fill}
\end{titlepage}

\section{Introduction}
In $\mathcal{N}=4$ supersymmetric Yang-Mills the Maldacena-Wilson loop operator for a single infinite straight line is a $\frac{1}{2}$-BPS object \cite{Maldacena:1998im, Drukker:1999zq}. As such it is believed that its vacuum expectation value does not receive any quantum corrections. Lower order perturbative calculations indicate that
\begin{align}\label{eq:VEV}
\big<\!\!\big< \mathcal{W}(-) \big>\!\!\big>_g = 1 \, .
\end{align}
However, thus far no rigorous proof exists showing that indeed all perturbative corrections cancel. In this work we compute the vacuum expectation value \eqref{eq:VEV} up to order $g^6$ for all $N$ using the recently fully established Nicolai map \cite{Ananth:2020gkt,Ananth:2020lup,Malcha:2021ess}, thus extending the previous perturbative results of Erickson, Semenoff and Zarembo \cite{Erickson:2000af, Zarembo:2002an} by one order. \\[2ex]
The importance of the infinite straight line Maldacena-Wilson loop can be seen in the context of the AdS/CFT correspondence. Supposing \eqref{eq:VEV} Drukker, Gross and Pestun have used the technique of localization to obtain the vacuum expectation value of a circular Maldacena-Wilson loop to all orders \cite{Drukker:2000rr,Pestun:2007rz}. Which in turn is related to the area of the minimal surface of a disk in supergravity \cite{Maldacena:1998im,Drukker:2000rr}. Thus providing a non-trivial test for the famous conjecture. \\[2ex]
In early work, Nicolai showed that supersymmetric gauge theories are characterized by a non-local and non-linear transformation $\mathcal{T}_g$, the Nicolai map \cite{Nicolai:1980jc, Nicolai:1982ye}. Then Dietz and Lechtenfeld realized that the Nicolai map provides a ghost and fermion free quantization of supersymmetric Yang-Mills theories \cite{Lechtenfeld:1984me, Dietz:1984hf, Dietz:1985hga}. However, only very recently it has been understood how to obtain the Nicolai map for all critical dimensions and in general gauges of $\mathcal{N}=1$ super Yang-Mills \cite{Ananth:2020gkt,Ananth:2020lup,Malcha:2021ess}. In \cite{Nicolai:2020tgo} Nicolai and Plefka used the map to compute some lower order examples of quantum correlators and the 1-loop dilation operator in $\mathcal{N}=4$ super Yang-Mills. \\[2ex]
Besides increasing the perturbative precision of \eqref{eq:VEV}, our calculation functions as a non-trivial proof of concept regarding the applicability of the Nicolai map. We show that our formalism is competitive with standard perturbative techniques. However, we will also see that the perturbative cancellations of the different contributions to the infinite straight line Maldacena-Wilson loop are by no means trivial and seem to resemble those of the circular Maldacena-Wilson loop at order $g^4$ (see \cite{Erickson:2000af}).

\subsection{The Maldacena-Wilson loop}
In $\mathcal{N}=4$ super Yang-Mills the Euclidean Maldacena-Wilson loop along a curve $\mathcal{C}$ is given by \cite{Maldacena:1998im} 
\begin{align}\label{eq:WilsonLoop}
\mathcal{W}^M(\mathcal{C}) = \frac{1}{N} \tr_c \, \mathcal{P} \, \exp \left( i g \int_{\mathcal{C}} \mathrm{d}\tau \left( A_\mu(x) \dot{x}^\mu + i \phi_I(x) |\dot{x}| \theta^I \right) \right) \, ,
\end{align}
where $A_\mu(x) = t^a A_\mu^a(x)$ is the gauge field and $\phi_I(x) = t^a \phi_I^a(x)$ are the six scalars. $\theta^I$ describes a point on the unit 5-sphere, \emph{i.e.} $\theta_I \theta^I = 1$, and $x(\tau)$ parametrizes the curve $\mathcal{C}$. In the following we choose the infinite straight line parametrized by $x^\mu(\tau) = (\tau, 0, 0, 0)$. The trace is over the color space with gauge group $U(N)$ in the fundamental representation and generators $t^a$ ($a = 1, \ldots , N^2$). These generators obey
\begin{align}
t^a t^a = \frac{N}{2} \mathbf{1} \, .
\end{align}
Furthermore, we have the relations
\begin{align}
f^{abc} f^{abd} = N \delta^{cd} \, , \quad \tr_c(t^a t^b) = \frac{1}{2} \delta^{ab} \, , \quad \tr_c( \mathbf{1} ) = N \, .
\end{align}
Starting at order $g^2$ the vacuum expectation value of a general Maldacena-Wilson loop is divergent when two or more space-time arguments approach each other. However, in \cite{Drukker:1999zq} it was argued that these linear divergences cancel for loops of the type \eqref{eq:WilsonLoop}, parametrized by a four vector $x^\mu$ and a point on the unit 5-sphere $\theta^I$. For an explicit proof at $\mathcal{O}(g^4)$ see \cite{Erickson:2000af}. In the case of the infinite straight line the situation is even simpler since up to $\mathcal{O}(g^4)$ all divergent terms are proportional to $\dot{x}_\mu \dot{x}^\mu - |\dot{x}| |\dot{x}| = 0$. However, we will see that at order $g^6$ this simplicity ceases to exist as the internal structure of in particular the 2- and 3-point correlation functions becomes more involved. We expect to obtain two UV divergent contributions from these correlation functions which cancel each other when they are summed up. \\[2ex]
Finally, partial contributions of for example fermion or ghost loops to any $n$-point function are in general highly divergent. Luckily the Nicolai map completely sidesteps the use of fermion and ghost fields in the computation of bosonic correlation functions and thus we will not see any divergences related to such loops.

\subsection{Correlation Functions and the Nicolai Map}
We state the main theorem from \cite{Nicolai:1980jc, Nicolai:1984jg,Ananth:2020lup}. \\[2ex]
\textit{Supersymmetric gauge theories are characterized by the existence of a non-linear and non-local transformation $\mathcal{T}_g$ of the bosonic fields $\Phi = \{A_\mu^a, \phi_I^a, \ldots \}$
\begin{align*}
\mathcal{T}_g \ : \ \Phi(x) \mapsto \Phi^\prime(x,g;\Phi) \, ,
\end{align*}
which is invertible at least in the sense of a formal power series such that
\begin{enumerate}
\item The bosonic action without gauge-fixing terms is mapped to the abelian action, 
\begin{align*}
S_B[g;\Phi] = S_B[0;\Phi^\prime] \, .
\end{align*}
\item The gauge-fixing function $\mathcal{G}^a(\Phi)$ is a fixed point of $\mathcal{T}_g$.
\item Modulo terms proportional to the gauge-fixing function $\mathcal{G}^a(\Phi)$, the Jacobi determinant of $\mathcal{T}_g$ is equal to the product of the MSS and FP determinants
\begin{align*}
\mathcal{J}(\mathcal{T}_g \Phi) = \Delta_{MSS}[\Phi] \Delta_{FP}[\Phi] \, , 
\end{align*}
at least order by order in perturbation theory. 
\end{enumerate}
} \noindent
The theorem was proven for $\mathcal{N}=1$ super Yang-Mills in $\mathcal{D}= 3,4,6$ and $10$ dimensions and Landau gauge in \cite{Ananth:2020lup} as well as in $\mathcal{D}=4$ dimensions and general gauges in \cite{Malcha:2021ess}. Furthermore, Rupprecht provided an extension to $\mathcal{N}=4$ super Yang-Mills \cite{Rupprecht:2021wdj} (also in Landau gauge). In \cite{Ananth:2020lup, Malcha:2021ess} the transformation $\mathcal{T}_g$ has been explicitly computed for $\mathcal{N}=1$ super Yang-Mills in $\mathcal{D}= 3,4,6$ and $10$ dimensions and Landau gauge up to the fourth order in the coupling. \\[2ex]
The inverse Nicolai map is obtained via the power series expansion of the operator $\mathcal{R}_g$ (see \emph{e.g.} \cite{Ananth:2020gkt})
\begin{align}
(\mathcal{T}_g^{-1} \Phi)(x) = \sum_{n=0}^\infty \frac{g^n}{n!} (\mathcal{R}_g^n \, \Phi)(x) \Big\vert_{g=0} \, .
\end{align}
Because the $\mathcal{R}_g$ operator has the properties of a derivative, the transformation $\mathcal{T}_g^{-1}$ acts distributively on bosonic monomials $X[\Phi]$, \emph{i.e.} 
\begin{align}
\mathcal{T}_g^{-1} X[\Phi] = X[\mathcal{T}_g^{-1} \Phi] \, .
\end{align}
The vacuum expectation value of such a bosonic monomial is given by
\begin{align}
\big<\!\!\big< X[\Phi] \big>\!\!\big>_g = \int \mathcal{D}\Phi \ \mathcal{D}\Psi \ e^{- S[g;\Phi,\Psi] } \ X[\Phi] \, ,
\end{align}
where $\Psi = \{ \lambda_\alpha^a, C^a, \ldots \}$ are the spinor and ghost fields and $S[g;\Phi,\Psi]$ is the full supersymmetric action including the gauge fixing and ghost terms of the theory in question. Integrating out the fermionic degrees of freedom gives
\begin{align}\label{eq:CorrX}
\big<\!\!\big< X[\Phi] \big>\!\!\big>_g = \int \mathcal{D}\Phi \ \Delta_{MSS}[\Phi] \Delta_{FP}[\Phi] \ e^{- S_B[g;\Phi]} \ X[\Phi] \, ,
\end{align}
where $S_B[g;\Phi]$ is the bosonic part of the supersymmetric action $S[g;\Phi,\Psi]$. The free field expectation value of $X[\Phi]$ is
\begin{align}
\big< X[\Phi] \big>_0 = \int \mathcal{D}\Phi \ e^{- S_B[0;\Phi]} \ X[\Phi] \, .
\end{align}
Considering the free field expectation value of $X[\mathcal{T}_g^{-1} \Phi]$ and performing a transformation of the integration variables yields
\begin{align}\label{eq:EVX}
\begin{aligned}
\big< X[\mathcal{T}_g^{-1} \Phi] \big>_0 &= \int \mathcal{D}\Phi \ e^{- S_B[0;\Phi]} \ X[\mathcal{T}_g^{-1} \Phi] \\
&= \int \mathcal{D}\Phi \ \mathcal{J}(\mathcal{T}_g \Phi) e^{- S_B[0;\mathcal{T}_g \Phi]} \ X[\Phi] \, .
\end{aligned}
\end{align}
Thus, if $\mathcal{T}_g$ satisfies \textit{1. - 3.} from the main theorem we find that \eqref{eq:CorrX} and \eqref{eq:EVX} are equal and we conclude
\begin{align}\label{eq:Monomial}
\big<\!\!\big< X[\Phi] \big>\!\!\big>_g = \big< X[\mathcal{T}_g^{-1} \Phi] \big>_0 \, .
\end{align} 
Notice that this transformation does not render the vacuum expectation value trivial, since the complexity is now hidden in the perturbative expansion of the non-linear and non-local transformation $\mathcal{T}_g^{-1}$. Using the linearity of $\big<\!\!\big<\ldots\big>\!\!\big>_g$ and the distributivity of $\mathcal{T}_g^{-1}$ we can extend \eqref{eq:Monomial} to $n$-point correlators of bosonic operators $\mathcal{O}_i(x_i)$, \emph{i.e.} 
\begin{align}\label{eq:CorrFct}
\big<\!\!\big< \mathcal{O}_1(x_1) \ldots \mathcal{O}_n(x_n) \big>\!\!\big>_g = \big< (\mathcal{T}_g^{-1}\mathcal{O}_1)(x_1) \ldots (\mathcal{T}_g^{-1}\mathcal{O}_n)(x_n) \big>_0 \, .
\end{align} 
So instead of computing full $n$-point correlation functions of the interacting super Yang-Mills theory (with fermions and ghosts), we can simply compute the free field expectation value of the purely bosonic non-interacting theory with the transformed operators. After working out the transformations $(T_g^{-1} \mathcal{O}_i)(x_i)$ to the desired order in the coupling, we simply use Wick's theorem to obtain the free field expectation value. Up to $\mathcal{O}(g^2)$ the inverse Nicolai map for $\mathcal{D}=10$, $\mathcal{N}=1$ super Yang-Mills in Landau gauge is given by
\begin{align}\label{eq:NicolaiMap}
\begin{aligned}
(\mathcal{T}_g^{-1} A)_M^a(z) &= A_M^a(z) - g f^{abc} \int \mathrm{d}v \ \partial^N C(z-v) A_M^b(v) A_N^c(v) \\
&\quad + \frac{g^2}{2} f^{abc} f^{bde} \int \mathrm{d}v \ \mathrm{d}w\ \Big\{ \\
&\quad \quad \quad + 3 \, \partial^N C(z-v) A^{c\, L}(v) \partial_{[M} C(v-w) A_N^d(w) A_{L]}^e(w) \\
&\quad \quad \quad - 4 \, \partial^N C(z-v) A_{[M}^c(v) \partial^L C(v-w) A_{N]}^d(w) A_{L}^e(w) \Big\} \\
&\quad + \mathcal{O}(g^3) \, .
\end{aligned}
\end{align}
This result was first found in \cite{Nicolai:1980jc} for $\mathcal{D}=4$, $\mathcal{N}=1$ super Yang-Mills. In \cite{Ananth:2020gkt} it was shown that it holds for all critical dimensions of $\mathcal{N}=1$ super Yang-Mills. So, in particular, in 10 dimensions.
\subsection{Conventions and Notation}
We use the Euclidean metric. Intermediate results in our calculations are UV divergent. Thus regularization by dimensional reduction is in order. For $\mathcal{N}=1$ super Yang-Mills in 10 dimensions we denote the spacetime indices by $M, N = 0, \ldots , 9$. Dimensionally reducing the 10-dimensional $\mathcal{N}=1$ theory to $\mathcal{N}=4$ super Yang-Mills in $2\omega$ dimensions, we split the spacetime indices $M = (\mu, I)$, where $\mu,\nu = 0, \ldots, 2\omega - 1$ and $I,J = 1, \ldots, 10 - 2\omega$. Likewise we decompose the coordinates $z^M = (x^\mu, y^I)$ and the gauge field
\begin{align}
A_M^a(x,y) = \big( A_\mu^a(x), \phi_I^a(x) \big) \, .
\end{align}
Notice that the dependence on the internal coordinates $y^I$ is dropped. \\[2ex]
The scalar propagator in $2\omega$ dimensions is (with the Laplacian $\Box \equiv \partial_\mu \partial^\mu$)
\begin{align}
C(x) = \int \frac{\mathrm{d}^{2\omega}k}{(2\pi)^{2\omega}} \frac{e^{ikx}}{k^2} \, .
\end{align}
It satisfies $- \Box \, C(x) = \delta(x)$ with the $2\omega$-dimensional delta function $\delta(x) \equiv \delta^{2\omega}(x)$. In $2\omega$ dimensions we have
\begin{align}
C(x) = \frac{\Gamma(\omega-1)}{4\pi^\omega} \frac{1}{[x^2]^{\omega-1}} \, .
\end{align}
In 10 dimensions the vector field propagator is
\begin{align}
\big< A_M^a(x) A_N^b(y) \big>_0 = \delta^{ab} \left( \delta_{MN} - (1-\xi) \frac{\partial_M \partial_N}{\Box} \right) C(x-y) \, .
\end{align}
Here $\xi$ is the gauge parameter. We argue that we can compute the inverse Nicolai map in Landau gauge ($\xi = 0$) whilst using the Feynman gauge ($\xi =1$) for the propagator because the Wilson loop is gauge invariant. So when computing its vacuum expectation value, all terms coming from the gauge parameter dependent term in the propagator must vanish. Thus without loss of generality we choose $\xi =1$ and the propagator becomes
\begin{align}\label{eq:AProp}
\big< A_M^a(x) A_N^b(y) \big>_0 = \delta^{ab} \delta_{MN} C(x-y) \, .
\end{align}
For $n$-point quantum correlation functions we define
\begin{align}
\big<\!\!\big< \mathcal{O}_1(x_1) \ldots \mathcal{O}_n(x_n) \big>\!\!\big>_m \coloneqq \big<\!\!\big< \mathcal{O}_1(x_1) \ldots \mathcal{O}_n(x_n) \big>\!\!\big>_g \bigg\vert_{\mathcal{O}(g^m)} \, , 
\end{align}
with $\big<\!\!\big< \mathcal{O}_1(x_1) \ldots \mathcal{O}_n(x_n) \big>\!\!\big>_0 = \big< \mathcal{O}_1(x_1) \ldots \mathcal{O}_n(x_n) \big>_0$. 

\section{Perturbation Theory}
In terms of the 10-dimensional fields the infinite straight line Wilson loop \eqref{eq:WilsonLoop} takes the simple form
\begin{align}\label{eq:VEV10D}
\mathcal{W}^M(-) = \frac{1}{N} \tr_c \, \mathcal{P} \, \exp \left( i g \int_{-\infty}^\infty \mathrm{d}\tau \ A_M(z) \dot{z}^M \right) \, ,
\end{align}
with the 10-dimensional gauge field $A_M(z) = t^a A_M^a(z)$ and $\dot{z}^M = ( \dot{x}^\mu, \dot{y}^I) = ( \dot{x}^\mu, i |\dot{x}| \theta^I)$. Moreover, we abbreviate $z_i \equiv z(\tau_i)$. For an infinite straight line $\dot{z}_i^M$ satisfies
\begin{align}
\delta_{MN} \,  \dot{z}_i^M \dot{z}_j^N = \dot{x}_i \cdot \dot{x}_j - |\dot{x}_i| |\dot{x}_j| = 0 \, .
\end{align}
 In perturbation theory the vacuum expectation value is given by
\begin{align}\label{eq:PertVEV}
\begin{aligned}
\big<\!\!\big< \mathcal{W}(-) \big>\!\!\big>_g = 1  &+ \frac{ig}{N} \int_{-\infty}^\infty \mathrm{d}\tau_1 \ \dot{z}_1^M \ \tr_c \, \big<\!\!\big< A_M(z_1) \big>\!\!\big>_g \\
& + \frac{i^2g^2}{2!N} \int_{-\infty}^\infty \mathrm{d}\tau_1 \ \mathrm{d}\tau_2 \ \dot{z}_1^M \dot{z}_2^N \ \tr_c \, \mathcal{P} \, \left<\!\!\left< A_M(z_1) A_N(z_2) \right>\!\!\right>_g  \\
& + \frac{i^3g^3}{3!N} \int_{-\infty}^\infty \mathrm{d}\tau_1 \ \mathrm{d}\tau_2 \ \mathrm{d}\tau_3 \ 
\dot{z}_1^M \dot{z}_2^N \dot{z}_3^L \ \tr_c \, \mathcal{P} \, \big<\!\!\big< A_M(z_1) A_N(z_2) A_L(z_3) \big>\!\!\big>_g \\
& + \ldots \, .
\end{aligned}
\end{align}
The expectation value has been computed perturbatively up to order $g^4N^2$ by Erickson, Semenoff and Zarembo in \cite{Erickson:2000af, Zarembo:2002an}. We have checked that their result also holds for all $N$. In the following we show how to compute the next nontrivial order of \eqref{eq:PertVEV} by the means of the Nicolai map. Expanding the vacuum expectation value at order $g^6$ we obtain
\begin{align}
\begin{aligned}\label{eq:W}
&\big<\!\!\big< \mathcal{W}(-) \big>\!\!\big>_6 \\
&\quad = \frac{ig}{N} \int_{-\infty}^\infty \mathrm{d}\tau_1 \ \dot{z}_1^M \ \tr_c \, \big<\!\!\big< A_M(z_1) \big>\!\!\big>_5 \\
&\quad \quad + \frac{i^2g^2}{2!N} \int_{-\infty}^\infty \mathrm{d}\tau_1 \ \mathrm{d}\tau_2 \ \dot{z}_1^M \dot{z}_2^N \ \tr_c \, \mathcal{P} \, \big<\!\!\big< A_M(z_1) A_N(z_2) \big>\!\!\big>_4 \\
&\quad \quad + \frac{i^3g^3}{3!N} \int_{-\infty}^\infty \mathrm{d}\tau_1 \ \mathrm{d}\tau_2 \ \mathrm{d}\tau_3 \ 
\dot{z}_1^M \dot{z}_2^N \dot{z}_3^L \ \tr_c \, \mathcal{P} \, \big<\!\!\big< A_M(z_1) A_N(z_2) A_L(z_3) \big>\!\!\big>_3 \\
&\quad \quad + \frac{i^4g^4}{4!N} \int_{-\infty}^\infty \mathrm{d}\tau_1 \ \mathrm{d}\tau_2 \ \mathrm{d}\tau_3 \ \mathrm{d}\tau_4 \ \dot{z}_1^M \dot{z}_2^N \dot{z}_3^L \dot{z}_4^P 
\ \tr_c \, \mathcal{P} \, \big<\!\!\big< A_M(z_1) A_N(z_2) A_L(z_3) A_P(z_4) \big>\!\!\big>_2 \\
&\quad \quad + \frac{i^5g^5}{5!N} \int_{-\infty}^\infty \mathrm{d}\tau_1 \ \mathrm{d}\tau_2 \ \mathrm{d}\tau_3 \ \mathrm{d}\tau_4 \ \mathrm{d}\tau_5 \\
&\quad \quad \quad \quad \times \dot{z}_1^M \dot{z}_2^N \dot{z}_3^L \dot{z}_4^P \dot{z}_5^Q 
\ \tr_c \, \mathcal{P} \, \big<\!\!\big< A_M(z_1) A_N(z_2) A_L(z_3) A_P(z_4) A_Q(z_5) \big>\!\!\big>_1 \\
&\quad \quad + \frac{i^6g^6}{6!N} \int_{-\infty}^\infty \mathrm{d}\tau_1 \ \mathrm{d}\tau_2 \ \mathrm{d}\tau_3 \ \mathrm{d}\tau_4 \ \mathrm{d}\tau_5 \ \mathrm{d}\tau_6 \\
&\quad \quad \quad \quad \times \dot{z}_1^M \dot{z}_2^N \dot{z}_3^L \dot{z}_4^P \dot{z}_5^Q \dot{z}_6^R 
\ \tr_c \, \mathcal{P} \, \big<\!\!\big< A_M(z_1) A_N(z_2) A_L(z_3) A_P(z_4) A_Q(z_5) A_R(z_6) \big>\!\!\big>_0 \, .
\end{aligned}
\end{align}
We briefly discuss the terms which vanish more or less trivially. The trace over 1-point function is zero since
\begin{align}
 \tr_c \, \big<\!\!\big< A_M(z_1) \big>\!\!\big>_5 = \tr_c(t^a) \, \big<\!\!\big< A_M^a(z_1) \big>\!\!\big>_5 = 0 \, .
\end{align}
For the 4-point function, we need to expand the inverse Nicolai map \eqref{eq:NicolaiMap} up to $\mathcal{O}(g^2)$. Then we use \eqref{eq:CorrFct} and collect all terms of $\mathcal{O}(g^2)$. Computing the Wick contractions, we obtain several non-vanishing terns. However, once we multiply the correlation function with $\dot{z}_1^M \dot{z}_2^N \dot{z}_3^L \dot{z}_4^P$ and insert the parametrization of the straight line, everything cancels. The vanishing of the last two terms is rather simple. In both cases, there are Wick contractions of two untransformed fields. These produce terms which are proportional to $\dot{z}_{M \, i} \dot{z}_j^M = 0$. Thus only the 2- and 3-point functions need to be discussed in detail.

\subsubsection*{2-point function}
In order to compute the 2-loop correction to the 2-point function we need to expand the inverse Nicolai map \eqref{eq:NicolaiMap} up to $\mathcal{O}(g^4)$\footnote{When computing the inverse map in Landau gauge it is necessary to explicitly enforce the gauge condition $\partial^\mu A_\mu = 0$ in all terms. This is similar to the determinant test for the Nicolai map in Landau gauge (see \cite{Ananth:2020lup}).}. For details see \cite{Ananth:2020lup} and Appendix B of \cite{Malcha:2021ess}, where $(\mathcal{T}_g \, A)_M^a$ up to order $g^4$ is given. When expanded to the fourth order $(\mathcal{T}_g^{-1} A)_M^a$ has about 500 terms. We apply \eqref{eq:CorrFct} to the 2-point function and collect all terms of $\mathcal{O}(g^4)$, \emph{i.e.}
\begin{align}
\begin{aligned}
\Sigma_1 &= \frac{i^2g^2}{2!N} \int_{-\infty}^\infty \mathrm{d}\tau_1 \ \mathrm{d}\tau_2 \ \dot{z}_1^M \dot{z}_2^N \ \tr_c \, \mathcal{P} \, \big<\!\!\big< A_M(z_1) A_N(z_2) \big>\!\!\big>_4 \\
 &= - \frac{g^6}{2N} \, \tr_c(t^a t^b) \int_{-\infty}^\infty \mathrm{d}\tau_1 \ \mathrm{d}\tau_2 \ \dot{z}_1^M \dot{z}_2^N \ \big< (\mathcal{T}_g^{-1} A)_M^a (z_1) (\mathcal{T}_g^{-1} A)_N^b (z_2) \big>_0 \ \Big\vert_{\mathcal{O}(g^4)} \, .
\end{aligned}
\end{align}
Because $\tr_c(t^a t^b) = \tr_c(t^b t^a)$ the path ordering is trivial. After computing the free field expectation value of the transformed fields and some basic simplifications, such as enforcing $f^{aab} = 0$, we obtain roughly 650 terms. Approximately a third of them are proportional to $\delta_{MN} \dot{z}_1^M \dot{z}_2^N = 0$. In order to reduce the number of remaining terms we observe that most of them are proportional to
\begin{align}
\begin{aligned}\label{eq:2PointExample}
&\int_{-\infty}^\infty \ \mathrm{d}\tau_1 \ \mathrm{d}\tau_2 \ \dot{z}_1^M \dot{z}_2^N  \int \mathrm{d}y_1 \ \mathrm{d}y_2 \ \mathrm{d}y_3 \ \mathrm{d}y_4 \ \\
&\quad \quad \times C(z_1 - y_1) \partial_M C(y_1-y_3) \partial^P C(y_1-y_4) C(y_3-y_4) \partial_P C(y_4 - y_2) \partial_N C(y_3-y_2) C(y_2-z_2) \, ,
\end{aligned}
\end{align}
where the four derivatives may sit at any of the seven propagators. Using integration by parts it is always possible to rearrange the contracted derivatives such that they act on two propagators both depending on either $y_1$, $y_3$ or $y_4$. In this situation we use
\begin{align}
\begin{aligned}
\int \mathrm{d}y_4 \ &\partial^P C(y_1-y_4) C(y_3-y_4) \partial_P C(y_4 - y_2) \\
&\quad \quad 
\begin{aligned}
 = \frac{1}{2} \int \mathrm{d}y_4 \ \Big\{ &- C(y_1-y_4) \Box \, C(y_3-y_4) C(y_4 - y_2) \\
&+ \Box \, C(y_1-y_4) C(y_3-y_4)  C(y_4 - y_2) \\
&+ C(y_1-y_4) C(y_3-y_4) \Box \, C(y_4 - y_2) \Big\} 
\end{aligned}
\end{aligned}
\end{align}
and $\Box \, C(x-y) = - \delta(x-y)$. Thus \eqref{eq:2PointExample} becomes
\begin{align}
\begin{aligned}
&\frac{1}{4} \int_{-\infty}^\infty \ \mathrm{d}\tau_1 \ \mathrm{d}\tau_2 \ \dot{z}_1^M \dot{z}_2^N \int \mathrm{d}y_1 \ \mathrm{d}y_2 \ \mathrm{d}y_3 \ \Big\{ \\
&\quad \quad \quad + \frac{1}{2} C(z_1 - y_1) \partial_M C(y_1-y_3)^2  \partial_N C(y_3-y_2)^2 C(y_2-z_2) \\
&\quad \quad \quad - C(z_1 - y_1) \partial_M C(y_1-y_3)^2  C(y_1 - y_2) \partial_N C(y_3-y_2) C(y_2-z_2) \\
&\quad \quad \quad - C(z_1 - y_1) \partial_M C(y_1-y_3)  C(y_1-y_2)   \partial_N C(y_3-y_2)^2 C(y_2-z_2) \Big\}  \, .
\end{aligned}
\end{align}
The first term turns out to be total derivative. Integrating it by parts we obtain
\begin{align}
\int_{-\infty}^\infty \mathrm{d}\tau_1 \ \dot{z}_1^M \partial_M C(z_1-y_1) \ [\ldots] = \int_{-\infty}^\infty \mathrm{d}\tau_1 \ \frac{\partial}{\partial \tau_1} \, C(z_1-y_1) \ [\ldots] = 0 \, .
\end{align}
The other two terms can be combined using the observation
\begin{align}
\int \mathrm{d}y_3 \ \partial_M C(y_1-y_3)^2 \partial_N C(y_3-y_2) = \int \mathrm{d}y_3 \ \partial_M C(y_1-y_3) \partial_N C(y_3-y_2)^2 \, .
\end{align}
We repeat these steps on the other 400 non-vanishing terms. Subsequently, we perform the dimensional reduction and obtain the now very simple expression
\begin{align}
\begin{aligned}
\Sigma_1 &= \frac{i^2g^2}{2!N} \int_{-\infty}^\infty \mathrm{d}\tau_1 \ \mathrm{d}\tau_2 \ \dot{z}_1^M \dot{z}_2^N \ \tr_c \, \mathcal{P} \, \big<\!\!\big< A_M(z_1) A_N(z_2) \big>\!\!\big>_4 \\
&= g^6 N^3 \int_{-\infty}^\infty \mathrm{d}\tau_1 \ \mathrm{d}\tau_2 \ \dot{x}_1^\mu \dot{x}_2^\nu \int \mathrm{d}y_1 \ \mathrm{d}y_2 \ \mathrm{d}y_3 \ \Big\{ \\
&\quad \quad \quad + \partial_\mu \partial_\nu C(x_1 - y_1) C(x_1-y_2) C(x_2 - y_1) C(y_1-y_3) C(y_2-y_3)^2 \\
&\quad \quad \quad + \frac{3}{2} \partial_\mu C(x_1-y_1) C(x_1-y_2) C(x_2-y_3) C(y_1-y_2) C(y_1-y_3) \partial_\nu C(y_2-y_3) \Big\} \, .
\end{aligned}
\end{align}
Neither of these two terms is a total derivative as there are two $x_1$ dependencies in each of them. Thus, we must cancel $\Sigma_1$ against the 3-point function.

\subsubsection*{3-point function}
For the 3-point function the procedure is much the same as for the 2-point function. For the trace and path ordering we find
\begin{align}
\begin{aligned}
\Sigma_2 &= \frac{i^3g^3}{3!N} \int_{-\infty}^\infty \mathrm{d}\tau_1 \ \mathrm{d}\tau_2 \ \mathrm{d}\tau_3 \ 
\dot{z}_1^M \dot{z}_2^N \dot{z}_3^L \ \tr_c \, \mathcal{P} \, \big<\!\!\big< A_M(z_1) A_N(z_2) A_L(z_3) \big>\!\!\big>_3 \\
&= - \frac{ig^3}{24N} \, d^{abc} \int_{-\infty}^\infty \mathrm{d}\tau_1 \ \mathrm{d}\tau_2 \ \mathrm{d}\tau_3 \ 
\dot{z}_1^M \dot{z}_2^N \dot{z}_3^L \, \big<\!\!\big< A_M^a(z_1) A_N^b(z_2) A_L^c(z_3) \big>\!\!\big>_3 \\
&\quad+ \frac{g^3}{24N} \, f^{abc} \int_{-\infty}^\infty \mathrm{d}\tau_1 \ \mathrm{d}\tau_2 \ \mathrm{d}\tau_3 \ \epsilon(\tau_1,\tau_2,\tau_3) \, 
\dot{z}_1^M \dot{z}_2^N \dot{z}_3^L \, \big<\!\!\big< A_M^a(z_1) A_N^b(z_2) A_L^c(z_3) \big>\!\!\big>_3 \, , 
\end{aligned}
\end{align}
where $d^{abc}$ is totally symmetric and
\begin{align}
\epsilon(\tau_1, \tau_2, \tau_3) = \left[ \theta(\tau_1 -\tau_2) - \theta(\tau_2 - \tau_1) \right] \left[ \theta(\tau_1 -\tau_3) - \theta(\tau_3 - \tau_1) \right] \left[ \theta(\tau_2 -\tau_3) - \theta(\tau_3- \tau_2) \right] \, .
\end{align}
So $\epsilon(\tau_1, \tau_2, \tau_3) = 1$ for $\tau_1 > \tau_2 > \tau_3$ and anti-symmetric under the transposition of any two $\tau_i$. The first term will cancel because the 3-point correlation function at $\mathcal{O}(g^3)$ is anti-symmetric in $a$, $b$, and $c$. This time we only need the inverse Nicolai map up to $\mathcal{O}(g^3)$. However, since we now compute a 3-point function instead of a 2-point function, after the Wick contraction, we have about the same number of terms as before. But two thirds of the terms are proportional to $\delta_{MN}$, $\delta_{ML}$ or $\delta_{NL}$ and thus cancel. The remaining terms are simplified using the same integration by parts relations as above. However, for the 3-point function there are no total derivatives. Subsequently, we perform the dimensional reduction and obtain the 15 terms
\begin{align}
\begin{aligned}
\Sigma_2 &= - \frac{g^6N^3}{12} \int_{-\infty}^\infty \mathrm{d}\tau_1 \ \mathrm{d}\tau_2 \ \mathrm{d}\tau_3 \ \epsilon(\tau_1,\tau_2,\tau_3) \ \dot{x}_1^\mu \dot{x}_2^\nu \dot{x}_3^\rho \int \mathrm{d}y_1 \ \mathrm{d}y_2 \ \mathrm{d}y_3 \ \Big\{ \\
&\quad \quad \quad + \partial_\mu \partial_\nu \partial_\lambda C(x_1-y_1) C(x_2-y_1) C(x_3-y_2) C(y_1-y_3) C(y_2-y_3)^2 \\
&\quad \quad \quad + \partial_\mu \partial_\nu C(x_1-y_1) \partial_\lambda C(x_2-y_1) C(x_3-y_2) C(y_1-y_3) C(y_2-y_3)^2 \\
&\quad \quad \quad + \text{permutations} \hspace*{223pt} \Big\} \\
&\quad + \frac{g^6N^3}{8} \int_{-\infty}^\infty \mathrm{d}\tau_1 \ \mathrm{d}\tau_2 \ \mathrm{d}\tau_3 \ \epsilon(\tau_1,\tau_2,\tau_3) \ \dot{x}_1^\mu \dot{x}_2^\nu \dot{x}_3^\rho \int \mathrm{d}y_1 \ \mathrm{d}y_2 \ \mathrm{d}y_3 \ \Big\{ \\
&\quad \quad \quad + \partial_\mu \partial_\nu C(x_1-y_1) C(x_2-y_2) C(x_3-y_3) C(y_1-y_2) C(y_1-y_3) \partial_\lambda C(y_2-y_3) \\
&\quad \quad \quad - C(x_1-y_1) \partial_\nu \partial_\lambda C(x_2-y_2) C(x_3-y_3) C(y_1-y_2) \partial_\mu C(y_1-y_3) C(y_2-y_3) \\
&\quad \quad \quad + C(x_1-y_1) C(x_2-y_2) \partial_\lambda \partial_\mu C(x_3-y_3) \partial_\nu C(y_1-y_2) C(y_1-y_3) C(y_2-y_3) \Big\} \, .
\end{aligned}
\end{align}
All these terms have a factor of the form
\begin{align}
\dot{x}_i^\mu \partial_\mu C(x_i-y) = \frac{\partial}{\partial \tau_i} \, C(x_i-y) 
\end{align}
and this is their only dependence on $x_i$. Thus we can integrate by parts and use
\begin{align}
\frac{\partial}{\partial \tau_1 } \epsilon(\tau_1,\tau_2,\tau_3) &= 2 \delta(\tau_1-\tau_2) - 2 \delta(\tau_1-\tau_3) \, .
\end{align}
After carrying out integrations over the delta functions and renaming the variables we obtain
\begin{align}
\begin{aligned}
\Sigma_2 &= g^6N^3 \int_{-\infty}^\infty \mathrm{d}\tau_1 \ \mathrm{d}\tau_2 \ \dot{x}_1^\mu \dot{x}_2^\nu \int \mathrm{d}y_1 \ \mathrm{d}y_2 \ \mathrm{d}y_3 \ \Big\{ \\
&\quad \quad \quad + \partial_\mu \partial_\nu C(x_1-y_1) C(x_1-y_1) C(x_2-y_2) C(y_1-y_3) C(y_2-y_3)^2 \\
&\quad \quad \quad - \partial_\mu \partial_\nu C(x_1-y_1) C(x_1-y_2) C(x_2-y_1) C(y_1-y_3) C(y_2-y_3)^2 \\
&\quad \quad \quad + \partial_\nu C(x_1-y_1) \partial_\mu C(x_1-y_1) C(x_2-y_2) C(y_1-y_3) C(y_2-y_3)^2 \\
&\quad \quad \quad - \partial_\mu C(x_1-y_1) C(x_1-y_2) \partial_\nu C(x_2-y_1) C(y_1-y_3) C(y_2-y_3)^2 \\
&\quad \quad \quad - \frac{3}{2} \partial_\mu C(x_1-y_1) C(x_1-y_2) C(x_2-y_3) C(y_1-y_2) C(y_1-y_3) \partial_\nu C(y_2-y_3) \Big\} \, .
\end{aligned}
\end{align}
The first and third term can be combined to give a total derivative. Also the fourth term is a total derivative. Subsequently, we conclude
\begin{align}
\begin{aligned}
\Sigma_2 &= -g^6N^3 \int_{-\infty}^\infty \mathrm{d}\tau_1 \ \mathrm{d}\tau_2 \ \dot{x}_1^\mu \dot{x}_2^\nu \int \mathrm{d}y_1 \ \mathrm{d}y_2 \ \mathrm{d}y_3 \ \Big\{ \\
&\quad \quad + \partial_\mu \partial_\nu C(x_1-y_1) C(x_2-y_1) C(x_1-y_2) C(y_1-y_3) C(y_2-y_3)^2 \\
&\quad \quad + \frac{3}{2} \partial_\mu C(x_1-y_1) C(x_1-y_2) C(x_2-y_3) C(y_1-y_2) C(y_1-y_3) \partial_\nu C(y_2-y_3) \Big\} \, .
\end{aligned}
\end{align}
We see that $\Sigma_1$ and $\Sigma_2$ cancel
\begin{align}
\Sigma_1 + \Sigma_2 = 0 \, .
\end{align}

\section{Conclusion}
We have shown that for a Maldacena-Wilson loop operator of an infinite straight line
\begin{align}
\big<\!\!\big< \mathcal{W}(-) \big>\!\!\big>_g = 1+ \mathcal{O}(g^8)
\end{align}
for all $N$. Despite the BPS nature of this operator, the cancellation of the perturbative corrections at the sixth order is far from trivial. They seem to resemble the cancellations of the fourth-order perturbative corrections for the expectation value of the circular Maldacena-Wilson loop (see \cite{Erickson:2000af}). All correlation functions have been computed using the Nicolai map. Despite the complexity of intermediate results, such as the non-linear and non-local transformation of the gauge field to fourth order in \cite{Malcha:2021ess}, the general procedure is rather simple as it completely circumvents the use of anti-commuting variables. In the future, it will be interesting to see if the Nicolai map can also be used to obtain non-perturbative results for certain Wilson loop operators. 

\section*{Acknowledgements}
I would like to thank H. Nicolai and J. Plefka for helpful discussions and H. Nicolai for reading the manuscript. The work of H.M. is supported by the IMPRS for Mathematical and Physical Aspects of Gravitation, Cosmology and Quantum Field Theory.

\end{document}